\documentclass{tMOP2e}

\usepackage{epstopdf}
\usepackage{widetext}
\usepackage{setspace}

\theoremstyle{plain}

\theoremstyle{definition}

\theoremstyle{remark}

\usepackage[breaklinks,colorlinks=true,citecolor={blue}, linkcolor= {blue}, urlcolor={blue}]{hyperref}

\begin{document}


\title{Competition between heating and cooling effects in an optomechanical oscillator using a squeezed field}

\author{
\name{Vinh N.T. Pham\textsuperscript{a}, 
Chu Manh Hoang\textsuperscript{b},
and Nguyen Duy Vy\textsuperscript{c,d,$^*$}\thanks{$^*$Corresponding author's email: nguyenduyvy@tdtu.edu.vn}}
\affil{
\textsuperscript{a}Department of Physics, Ho Chi Minh City University of Education, Ho Chi Minh City, Vietnam;
\textsuperscript{b}International Training Institute for Materials Science, Hanoi University of Science and
Technology, Hanoi 10000, Vietnam;
\textsuperscript{c}Laboratory of Applied Physics, Advanced Institute of Materials Science, Ton Duc Thang University, Ho Chi Minh City, Vietnam; 
\textsuperscript{d}Faculty of Applied Sciences, Ton Duc Thang University, Ho Chi Minh City, Vietnam;
}
}
\maketitle

\setstretch{1.6}

\begin{abstract}
Squeezed light is a useful phenomenon that can be exploited to improve the sensitivity of specific classes of detectors based on optomechanical effects. Recently, there has been significant interest in the potential application of a squeezed field in the cooling of an optomechanical oscillator. It has been shown that this field could cool an oscillator below the standard limit of a coherent field. In this study, the effect of squeezed light was evaluated by explicitly examining the role of the squeezing parameters on the final effective temperature of the oscillator. The results show that the observed cooling and heating effects are strongly dependent on the squeezing parameters and the phase. Using an oscillator of 2$\pi\times$10.1 MHz driven by a 1064-nm laser, the lowest effective temperature and quantum number are three orders of magnitude smaller compared to the case of no squeezing; especially, these minimum values are obtained at the squeezing phase of about 0.8$\pi$. This study highlighted important insights for the optimization of cooling efficiency using squeezed light.
\end{abstract}


\section{Introduction}
The optomechanical oscillator is an essential element in various fields of research such as the measurement of gravitational waves \cite{CorbittPRA06,TofighiOptC10} and quantum information technologies \cite{
Rabl_nphys,YiOptC15}
. As a tested mass, when downscaled closer to the quantum regime, the effects of noise inevitably increase and obscure the observation results. Therefore, a systematic study of noise reduction is crucial. The dynamics of a tested mass exhibit a strong dependence on optomechanical coupling, in which a radiation force (RF) plays the role of the driving force \cite{CarmonPRL05, Hoffmann,BahrampourOptC11,KumarOptC12}. Such a coupling modifies the mass's motion and gives rise to damping via the back action of photon collision and photothermal induction \cite{KarraiNat, OkamotoPRL11}. This effect can be strongly enhanced in microcavity systems with high finesse because the effective elastic rigidity is modulated by the stored photons
. At the near resonance position, the intracavity power has a quadratic dependence on small changes in the length of the cavity, and a linear dependence in the case of a detuned cavity. To achieve appropriate detuning, it was reported that passive optical cooling \cite{KarraiNat} in a micro-mechanical resonator quenched fluctuation of a gold-coated silicon cantilever to an effective temperature of 18 K using light-induced force. This is the result of light scattering and absorption that is enhanced in the cavity \cite{GenesPRAcool}. Recent calculations \cite{OptLett02, VyAPL13, VyAPEX15,VyAPL16} using Maxwell's stress tensor has shown a Lorentzian form of the RF inside the cavity. In the vicinity of the resonance position, a linear dependence of the RF on the displacement could be implemented, and a linear optomechanical coupling strength is adopted.

Instead of using coherent light as the driving field to control and cool the dynamics of a mechanical oscillator, recent studies have investigated squeezed light as a potential driving source. 
Squeezed light could be created by parametric amplification \cite{DungOpt97,
SchnabelOptC04squeez} or by using radiation pressure shot noise \cite{PurdyPRX13squeez}. This phenomenon was first observed \cite{WallNature83} in a four-wave mixing experiment using a parametric oscillator or a parametric down-converter. 
With quantum fluctuations below that of a vacuum field, this light could be used in many high precision measurements because it could reduce the optical read-out noise \cite{SafaviNat13squeeze}, monitor mechanical motion, enhance the feedback control of the mechanical mode \cite{SchaferNatCom16,LotfipourPRA16squeez}, and elucidate the classical-quantum boundary \cite{BaiSciRep17,YinPRA17}.
The potential role of squeezed light in the cooling of mechanical oscillators has generated significant interest from both theoretical and experimental viewpoints \cite{AsjadPRA16,ClarkNat17}. Several effects are expected to occur at a very low noise level once the oscillator is cooled to sub-millikelvin temperatures. However, a detailed investigation of the dependence of the cooling efficiency on the squeezing parameters and phase has not been performed.

In this study, squeezed light is used as the driving field for a mechanical oscillator. By explicitly examining the role of the squeezing parameters and phase on the final mechanical energy of the oscillator, we show that the oscillator could be heated or cooled depending on the characteristics of the driving field. For a mechanical oscillator of $\omega_m$ = 2$\pi\times$10.1 MHz driven by a ($\lambda$ =) 1064-nm (Nd:YAG) laser, the minimal effective temperature and quantum number were seen at the squeezing phase of 0.8$\pi$. It is worth to noting that this phase is the result of the competition between the heating and cooling effects and is dependent on the parameters of the system, e.g. the mass and frequency of the oscillator ($m$ and $\omega_m$) and the optical wavelength ($\lambda$). 
The effective temperature and quantum number of the oscillator could be three orders of magnitude lower compared to non-squeezing light. This reduction is much lower than the value obtained in previous studies whereby after the optimization of all parameters, a decrease in temperature of approximately one to two orders of magnitude was reported. 


\section{Hamiltonian formalism for optomechanical systems}\label{Hamiltonian}
\begin{figure}[!h] \centering
\includegraphics[width=0.5\textwidth]{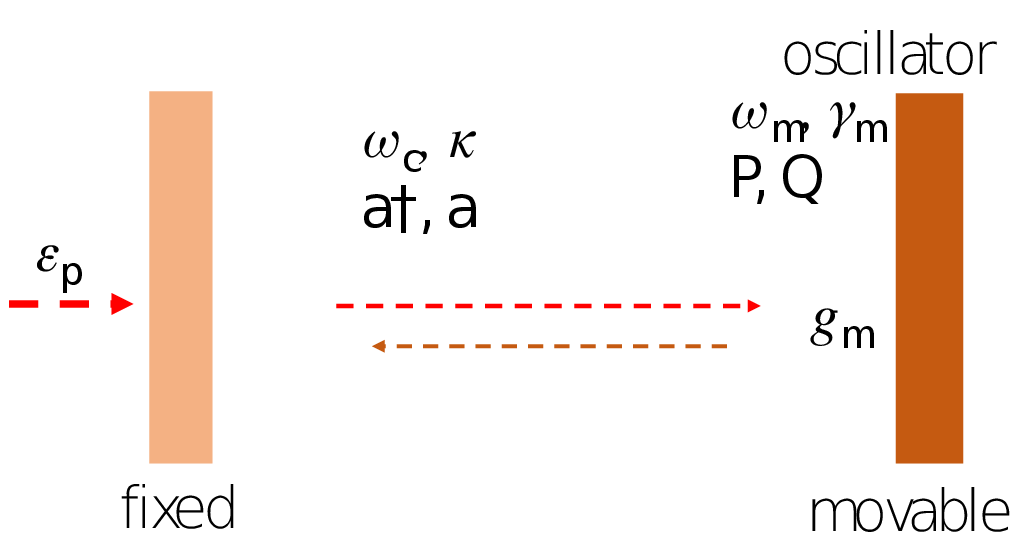}
\caption[Optical microcavity in Hamiltonian formalism.]{Optical microcavity in Hamiltonian formalism wherein the cavity photons (frequency $\omega_c$ and damping rate $\kappa$) couple with the mechanical mode (frequency $\omega_m$ and damping rate $\gamma_m$) depending on the coupling strength $g_m$.} \label{fig.OMCqm}
\end{figure}

Consider an optical cavity system with one movable mirror that can be considered to be a mechanical oscillator, as shown in Fig. \ref{fig.OMCqm}. 
The fixed mirror is assumed to be semi-transparent and the movable mirror is totally reflective. When the system is irradiated by a laser, the trapped light between the mirrors could be significantly enhanced and the intensity is dependent on the position of the movable mirror near the resonance position. The momentum of the photons creates a radiation pressure (RP) on the mirror, which could displace the oscillator out of its equilibrium position. 
Once the oscillator is displaced, the system loses its resonance, i.e., the cavity length is no longer a multiple of half laser wavelengths ($\lambda=\omega_p$/c where $\omega_p$ is the laser frequency and c is the speed of light) and the light intensity inside the cavity is reduced, in addition to the RP exerted on the oscillator. This means the oscillator could return its previous equilibrium position because the pushing force is reduced and the Hooke's force is increased. This is the coupling mechanism between the laser and the mechanical oscillator.  

In the Hamiltonian formalism, the optical modes are quantized and the mechanical one is kept as a \textit{c}-number. 
The Hamiltonian of the oscillator is $H_m =(1/2)(P^2/m+m\omega_m^2Q^2)$ where $P$ is the momentum operator and $Q$ is the position operator that satisfies $[P, Q] =-i\hbar$. 
Using $p=P/\sqrt{m\hbar\omega_m}$ and $q=Q\sqrt{m\omega_m/\hbar}$, we obtain $H_m =(1/2)\hbar\omega_m(p^2+q^2)$, where the $q$ and $p$ operators satisfy the commutator $[q, p]=i$. 

The linear coupling (interaction) between the optical and the mechanical mode is written as 
\begin{align}
H_i=-\hbar a^\dagger a\frac{\partial \omega_c}{\partial x}\delta x= -\hbar a^\dagger a\Big(\frac{\partial \omega_c}{\partial x}x_{ZPF}\Big)q =-\hbar g_ma^\dagger a q, 
\end{align}
where $g_m =(\partial\omega_c/\partial x)x_{ZPF} =(\partial\omega_c/\partial x)\sqrt{\hbar/2\omega_m m}$ is the coupling strength. $\partial\omega_c/\partial x$ is usually set to be $\omega_c/L_c$ in the literature, \cite{LawPRA2537} therefore $g_m =(\omega_c/L_c) x_{ZPF}$. 
The total Hamiltonian in the rotating frame \cite{WRaePRL07} of the pump laser with frequency $\omega_p$ [with power $P_i$, amplitude $\epsilon_p=[P_i\kappa/(2\hbar\omega_p)]^{1/2}$, and the cavity damping rate 2$\kappa$] is
\begin{align}
H=\hbar\Delta_0 a^\dagger a +\frac{1}{2}\hbar\omega_m(p^2+q^2) -\hbar g_ma^\dagger aq +i\hbar\epsilon_p(a^\dagger-a), 
\end{align}
 where $\Delta_0 = \omega_c-\omega_p$ is the cavity-laser detuning and $\omega_c=2\pi c/L_c$ is the frequency of the cavity mode. 
Using the Heisenberg equations for these operators
and adding noise operators and damping, the following Langevin equations are obtained,
\begin{subequations} \label{eq.Langevin}
\begin{align}
\dot{a}&= -(\kappa+i\Delta_0)a+ig_mqa +\epsilon_p +\sqrt{2\kappa}a_{in}, \label{dot_a}\\
\dot{a}^\dagger &= -(\kappa-i\Delta_0) a^\dagger -i g_m a^\dagger q +\epsilon_p +\sqrt{2\kappa}a_{in}^\dagger, \label{dot_adag}\\
\dot{p}&= -\gamma_m p -\omega_m q +g_m a^\dagger a +\xi(t), \label{dot_p}\\
\dot{q}&= \omega_mp,
\end{align}\end{subequations}
 where $2\gamma_m$ is the mechanical damping rate. The cavity mode is affected by the vacuum radiation
input noise $a_{in}$ and follows the correlation functions \cite{walls2008quantum,HuangNJP09}
\begin{subequations}\label{corre.phot}\begin{align}
\langle a_{in}(t)a_{in}^\dagger(t')\rangle &=[N(\omega_c)+1]\delta(t-t'),\\
\langle a_{in}^\dagger(t)a_{in}(t')\rangle &=N(\omega_c)\delta(t-t'),\\
\langle a_{in}(t) a_{in}(t')\rangle &=M(\omega_c)e^{-i\omega_m(t+t)}\delta(t-t'),\\
\langle a_{in}^\dagger(t) a_{in}^\dagger(t')\rangle &=M^*(\omega_c)e^{i'\omega_m(t+t)}\delta(t-t'),
\end{align}\end{subequations}
where 
$M$ is the two-photon correlation function and $\phi$ is the phase of the squeezed field. $N=\sinh^2r$ and $M=\sinh r \cosh r e^{i\phi}$ where $r$ is the squeezing parameter and $\phi$ the squeezing phase \cite{HuangNJP09}.
The fluctuation of the mechanical mode is \cite{GioavaPRA01noise, NewJPdyn, walls2008quantum}
\begin{subequations}\label{corre.fre}\begin{align}
\langle a_{in}(\omega)a_{in}^\dagger(\omega')\rangle &=\delta(\omega-\omega'), \label{corre.phot.fre}\\
\langle \xi(\omega)\xi(\omega')\rangle & =\frac{\gamma_m}{\omega_m}\omega\big[\coth(\frac{\hbar\omega}{2k_BT})+1\big]\delta(\omega+\omega'), \label{corre.phon.fre} 
\end{align}\end{subequations}
and all other correlators are zero.
The steady state solutions are as follows:
\begin{eqnarray} \label{steady}
\begin{cases} a_s =& \dfrac{\epsilon_p}{\kappa +i(\Delta_0 -g_mq_s) } \\
a_s^\dagger =& \dfrac{\epsilon_p}{\kappa -i(\Delta_0 -g_mq_s) }\\
q_s =& g_m|a_s|^2/\omega_m \\
p_s =&0\end{cases}.
\end{eqnarray} 
The fluctuation spectra of the transmitted field are examined by linearizing the quantum Langevin equation. The operators are written as the summation of their mean values and the fluctuation operators \cite{GenesPRAcool, VitaliPRLmirrorField, VitaliA84phasenoise}, $a=a_s+\delta a$, $p= p_s +\delta p$, $q= q_s +\delta q$. Thus we have:
\begin{subequations}\label{dot.equ1}\begin{align}
\delta\dot{a} 
 =& -(\kappa+i\Delta)\delta a +iG_a\delta q +\sqrt{2\kappa}a_{in},\\
\delta \dot{a}^\dagger =& -(\kappa-i\Delta)\delta a^\dagger -iG_a^*\delta q +\sqrt{2\kappa}a_{in}^\dagger,\\
\delta\dot{p} 
 =& -\gamma_m\delta p -\omega_m\delta q +G_a^*\delta a+G_a\delta a^\dagger +\xi ,\\
\delta\dot{q}=& \omega_m\delta p,
\end{align}\end{subequations} where $\Delta=\Delta_0-g_mq_s$ and $G_a = g_m a_s$. We rewrite Eq. (\ref{dot.equ1}) as
\begin{widetext}
\begin{eqnarray} \label{sys_eqs}
\left( \begin{array}{c} 
\delta \dot{a} \\ \delta \dot{a}^\dagger \\\delta \dot{p} \\\delta \dot{q} \end{array} \right) =
 \begin{pmatrix} 
 -(\kappa+i\Delta) & 0 & 0 & iG_a \\ 
0 & -(\kappa -i\Delta) & 0 & -iG_a^* \\ G_a^* & G_a &-\gamma_m & -\omega_m \\ 0&0& \omega_m &0 \end{pmatrix} 
 \left( \begin{array}{c} \delta a \\ \delta a^\dagger \\\delta p \\\delta q \end{array} \right) + 
 \left( \begin{array}{c} \sqrt{2\kappa}a_{in} \\ \sqrt{2\kappa}a_{in}^\dagger \\ \xi \\ 0 \end{array} \right).
\end{eqnarray} 
Taking the Fourier transform, we have: 
 $\mathcal{F}[\delta \dot{a}(t)] \rightarrow -i\omega\delta a(\omega)$, the above matrix then becomes
\begin{eqnarray}
 \begin{pmatrix} 
 -i\omega+\kappa+i\Delta & 0 & 0 & -iG_a \\ 
 0 & -i\omega+\kappa -i\Delta & 0 & iG_a^* \\ 
 - G_a^* & - G_a & -i\omega+\gamma_m & \omega_m \\ 
 0&0& -\omega_m & -i\omega \end{pmatrix} \left( \begin{array}{c} \delta a \\ \delta a^\dagger \\\delta p \\\delta q \end{array} \right) = 
 \left( \begin{array}{c} \sqrt{2\kappa}a_{in} \\ \sqrt{2\kappa}a_{in}^\dagger \\ \xi \\ 0 \end{array} \right). \label{eqFourier}
\end{eqnarray} 
\end{widetext} 
As a linear system of equations, Eq. (\ref{eqFourier}) has solutions if some conditions are satisfied i.e., the Routh-Hurwitz criterion for the parameters to be satisfied. This limits the input power and other mechanical parameters \cite{VyAPEX15}. 
Considering the steady states in Eq. (\ref{steady}), we can choose the relative phase reference for the intracavity field and the external laser so that $a_s$ is real and positive, for example, $$\epsilon_p =|\epsilon|e^{-i\theta} =|\epsilon|\frac{\kappa+i(\Delta_0-g_mq_s)}{\sqrt{\kappa^2+(\Delta_0-g_mq_s)^2}},$$ we denote $G^*_a = G_a =G$. Assuming that the Routh-Hurwitz criterion is satisfied \cite{RouthCriterion}, the solution of Eq. (\ref{eqFourier}) is
\begin{widetext}
\begin{align}
\delta q(\omega) =& \frac{-\omega_m}{d(\omega)}
\Big\{
\big[\Delta^2+(\kappa-i\omega)^2\big] \xi -i G\sqrt{2\kappa}\big[(\omega+i\kappa-\Delta)a_{in}^\dagger +(\omega+i\kappa+\Delta)a_{in} \big]
\Big\}, \label{delq}
\\
\delta p(\omega) =& \frac{-i\omega}{\omega_m} \delta q(\omega), \label{delp}\\
\delta a(\omega) =& \frac{-1}{d(\omega)}
\Big\{
G\omega_m(\omega+i\kappa+\Delta)\xi +i G^2\omega_m\sqrt{2\kappa}a_{in}^\dagger + \nonumber\\ 
&
-i\big[(\omega+i\kappa+\Delta)(\omega^2-\omega_m^2+i\gamma_m\omega)-G^2\omega_m\big]\sqrt{2\kappa}a_{in}
\Big\}, \label{dela}
\\
\delta a^\dagger(\omega) =& \frac{-1}{d(\omega)}
\Big\{
G\omega_m(\omega+i\kappa-\Delta)\xi -i G^2\omega_m\sqrt{2\kappa}a_{in}
+\nonumber\\ &
+i\big[(\omega+i\kappa-\Delta)(\omega^2-\omega_m^2+i\gamma_m\omega)-G^2\omega_m\big]\sqrt{2\kappa}a_{in}^\dagger
\Big\}. \label{delap}
\end{align}
\end{widetext}
where 
\begin{align}
d(\omega) =2\Delta G^2\omega_m +(\omega+i\kappa-\Delta)(\omega+i\kappa+\Delta)(\omega_m^2-\omega^2-i\omega\gamma_m).
\end{align}

\section{Phonon spectrum}\label{sec.phonon}
We rewrite Eq. (\ref{delq}) as
\begin{align}
\delta q(\omega) &= \frac{-\omega_m}{d_s(\omega)}
\Big\{ \xi 
+i G\sqrt{2\kappa}\big[\frac{a_{in}^\dagger}{\omega+i\kappa+\Delta} +\frac{a_{in}}{\omega+i\kappa-\Delta} \big]
\Big\}, \label{delqr}
\end{align}
where 
\begin{align}
d_s(\omega) =\frac{d(\omega)}{\Delta^2+(\kappa-i\omega)^2} = \omega_{eff}^2-\omega^2-i\omega\gamma_{eff}, \label{ds.re.ana}
\end{align}
with the effective resonance frequency, $\omega_{eff}$, and the effective damping rate, $\gamma_{eff}$,  
\begin{subequations}\label{omgam.effGe}
\begin{align}
\omega_{eff}^2(\omega) &=\omega_m^2+G^2\omega_m\frac{2\Delta (\omega^2-\Delta^2-\kappa^2)}{[(\omega-\Delta)^2+\kappa^2](\omega+\Delta)^2+\kappa^2]},\label{om.effGe}\\
\gamma_{eff}(\omega) &=\gamma_m+G^2\omega_m\kappa\frac{4\Delta}{[(\omega-\Delta)^2+\kappa^2](\omega+\Delta)^2+\kappa^2]}. \label{gam.effGe}
\end{align}
\end{subequations}
The phonon spectrum, $S_q$, is obtained from the phonon variance $\delta q(\omega)$ in Eq. (\ref{delqr}), $S_q(\omega)=
\langle\delta q(\omega)\delta q^*(\omega)\rangle$
. We have
\begin{widetext}
 \begin{align}
\langle\delta q\delta q^*\rangle =& \frac{\omega_m^2}{|d_s(\omega)|^2} 
\Big\{
\langle\xi\xi\rangle +G^22\kappa
\Big[
\frac{1}{\omega+i\kappa-\Delta}\frac{1}{\omega-i\kappa-\Delta}\langle a_{in}a_{in}^\dagger\rangle 
+\frac{1}{\omega+i\kappa+\Delta}\frac{1}{\omega-i\kappa+\Delta}\langle a_{in}^\dagger a_{in}\rangle
+\nonumber\\
& +\frac{1}{\omega+i\kappa+\Delta}\frac{1}{\omega-i\kappa-\Delta}\langle a_{in}^\dagger a_{in}^\dagger\rangle
+\frac{1}{\omega+i\kappa-\Delta}\frac{1}{\omega-i\kappa+\Delta}\langle a_{in} a_{in}\rangle
\Big]
\Big\} \label{delqdelq_0}
\\
=&\frac{\omega_m^2}{|d_s(\omega)|^2} 
\Big\{
\frac{\gamma_m\omega}{\omega_m}\big[\coth(\frac{\hbar\omega}{2k_BT})+1\big] +\frac{G^22\kappa}{(\omega-\Delta)^2+\kappa^2}
\Big\} +
\delta S_N+\delta S_M, \label{delqdelq}
\end{align}
\end{widetext}
where 
\begin{align}
\delta S_N =&\frac{\omega_m^2}{|d_s(\omega)|^2} 
\Big\{ 2G^2\kappa{ {N}}(\omega_c)
\Big[
\frac{1}{(\omega-\Delta)^2+\kappa^2} +\frac{1}{(\omega+\Delta)^2+\kappa^2}
\Big] \Big\},
\label{eq_SN}
\end{align}
and $S_M$ could be split as
\begin{align}
\delta S_M 
=&\frac{\omega_m^2}{|d_s(\omega)|^2} 
\Big\{
4G^2\kappa M(\omega)
\frac{(\omega^2-\Delta^2+\kappa^2)\cos\phi -2\kappa\Delta\sin\phi}{[(\omega+\Delta)^2+\kappa^2][(\omega-\Delta)^2+\kappa^2]}
\Big\}.
\label{eq_SM}
\end{align}

The total mechanical energy of the oscillator is $E_m=\frac{1}{2}\hbar\omega_m(\langle q^2\rangle +\langle p^2\rangle)$, where 
$\langle q^2\rangle = \frac{1}{2\pi}\int_{-\infty}^\infty S_q(\omega) d\omega,
$ and similar for $\langle p^2\rangle$. In optomechanics, the effective temperature ($T_{eff}$) is also used to present this energy via the fluctuation-dissipation theorem, $E_m=\frac{1}{2}k_BT_{eff}$ ($k_B$ is the Boltzmann constant). 
Finally, 
the oscillator energy is
\begin{align}
E_m=E_{th} +N(\omega)E_N + M(\omega)(E_M^c \cos\phi -E_M^s \sin\phi). \label{eq_Em}
\end{align}
where $E_{th}$, from the first term in Eq. (\ref{delqdelq}), is the energy that is in equilibrium with the thermal bath at a temperature $T$. $E_N$, from Eq. (\ref{eq_SN}), and $E_M^c$ and $E_M^s$, from Eq. (\ref{eq_SM}) are due to the coupling with the squeezed field. The integrations are independent on the squeezing phase, $\phi$. 
Based on the preceding analysis, we can summarize the following points:
\begin{itemize}
\item $E_{th}$ is the phonon energy under coupling with the vacuum or with a coherent field. This term is independent on the squeezing parameters and can be obtained by using a coherent driving field. Increasing the optomechanical coupling strength  [see Fig. \ref{fig_Sq}, $G$ = 0.15$\omega_m$ (black solid line) to 0.3$\omega_m$ (wine dash-dotted line)] shifts and suppresses the mechanical spectral function and results in cooling, as seen in recent studies \cite{GenesPRAcool,VyAPEX16}
\begin{figure}[!ht] \centering
\includegraphics[width=0.65\textwidth]{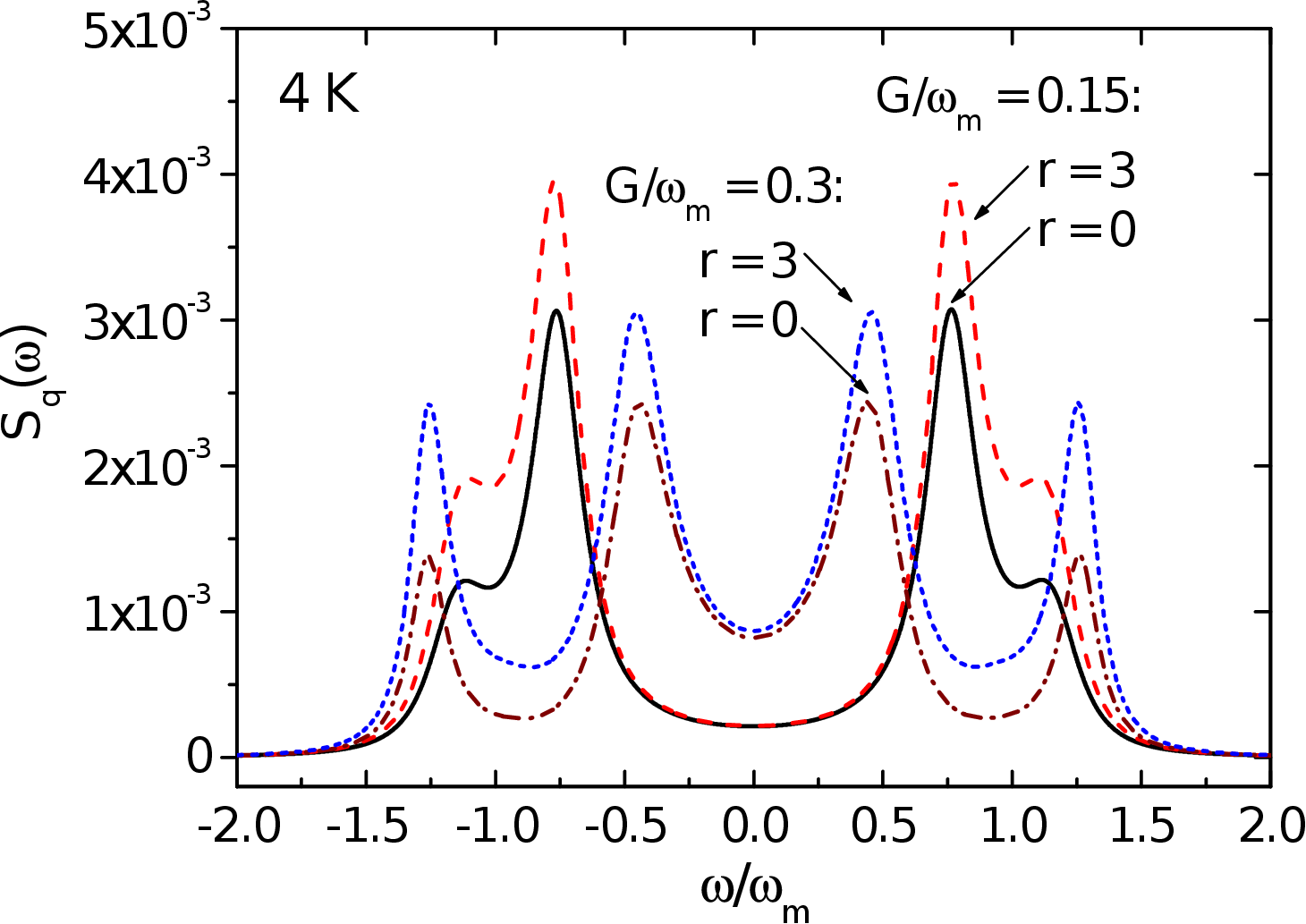}
\caption{Spectra $S_q(\omega)$of the mechanical modes for increasing the optomechanical coupling strength ($G$). $S_q(\omega)$ is suppressed and the peak at $\omega_m$ is usually split and shifted to smaller values.} \label{fig_Sq}
\end{figure} 
\item $N(\omega) E_N$ appears due to the correlation between the two annihilation or two creation operators, $N(\omega)\propto \langle a a\rangle$ and $\langle a^\dagger a^\dagger\rangle$, as shown in Eq. (\ref{corre.phot}). This term represents the characteristics of the squeezed vacuum and diminishes in a coherent state, $N(\omega)\simeq$ 0. However, in the squeezed field, it has a significant contribution to the phonon energy.
\item If the squeezing parameters are not suitably chosen, the contributions due to $N(\omega)$ and $M(\omega)$, Eqs. (\ref{delqdelq})--(\ref{eq_SM}) can be positive and the spectral function is enhanced, as shown in Fig. \ref{fig_Sq} (red dashed line and blue dotted line) for $\phi$ =0.
\item $E_M^c$ and $E_M^s$ exhibit a significant change in their magnitudes versus the optomechanical coupling strength. As shown in Fig. \ref{fig_Er}, $E_M^c > E_M^s$ and both increase with the coupling strength $G$. 
Furthermore, they are strongly enhanced with increasing $r$, from smaller than [see Fig. (a)] to greater than $E_{th}$ [see Fig. (c)].
\begin{figure*}[!ht] \centering
\includegraphics[width=1.01\textwidth]{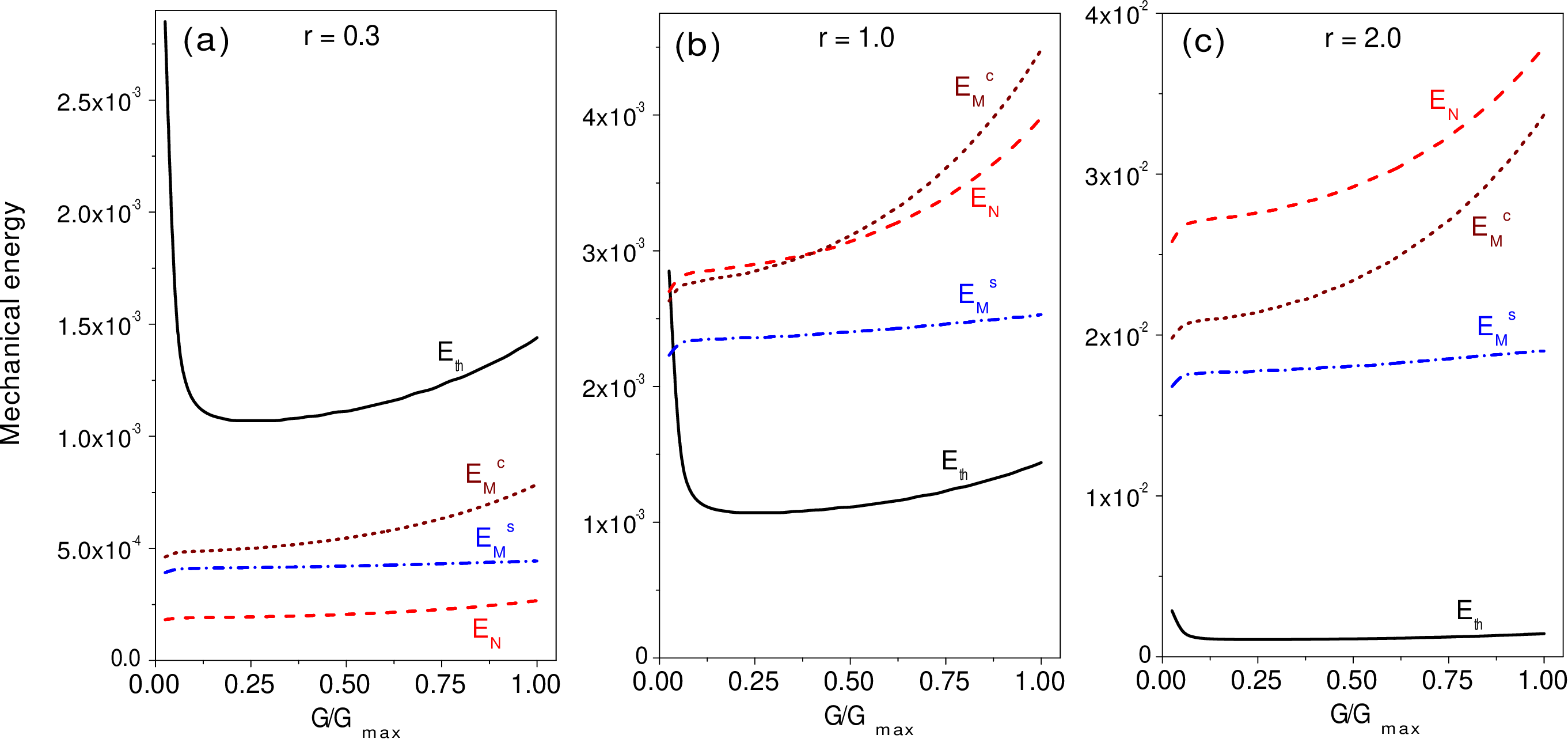}
\caption{Mechanical energy the mechanical mode from separate contributions that are presented in Eq. (\ref{eq_Em}). $E_N$, $E_M^c$, and $E_M^s$ increase with the increase of the squeezing parameter ($r$), whereas $E_{th}$ is independent of $r$. These energies exhibit a significant change of their values with increasing coupling strength $G$. In this case, $\omega_m = 2\pi\times$10.1 MHz, $\gamma_m = 2\pi\times$150 Hz, $g_m = 2\pi\times$260 Hz, and $T$ = 37 mK.} \label{fig_Er}
\end{figure*} 
\item The contribution of $E_N$ is always positive. However, that of $E_M^c$ and $E_M^s$ with their corresponding $\cos$ and $\sin$ could be positive or negative. As a result, the total contribution is strongly dependent on the squeezing phase $\phi$ and the coupling strength $G$. 
\end{itemize}

\section{Competition between heating and cooling}
By choosing a suitable squeezing parameter $r$, the summation due to $E_N$ and $E_M^{c,s}$ could be positive or negative. When this summation is positive, the squeezed field heats the oscillator. Otherwise, the squeezed field cools the oscillator to an effective temperature below the coherent value $E_{th}$. 
If the contributions of $E_N$ and $E_M^{c,s}$ are known, we can determine the optimal parameters of the squeezed field so that the oscillator is cooled beyond the case of a coherent field. To achieve a negative contribution, the total energy exchanged by the squeezed field,
\begin{align}\Delta E= N(\omega)E_N +M(\omega)(E_M^c\cos\phi-E_M^s\sin\phi). \label{eq_delE}\end{align}
must be negative. 
We could see from Eq. (\ref{delqdelq}) that 
 $S_q(\omega) \propto\frac{\omega_m^2}{|d_s(\omega)|^2} \{\xi(\omega)\}^2$ where $\xi(\omega)$ in $\{$...$\}$ plays the role of external noise sources. 
If the susceptibility function $\chi(\omega)=\frac{\omega_m^2}{|d_s(\omega)|^2}$ is diminished, the oscillator has a weaker response to the thermal bath and its energy is also reduced. Therefore, enhancement of $d_s$ by changing $\omega_{eff}$ and $\gamma_{eff}$ is the key to optomechanical cooling. The minimum effective temperature obtained using this method is strongly limited by the mechanical structure of the oscillator. In particular, the squeezed vacuum, whose properties can be effectively controlled by changing the squeezed parameter $r$ and phase $\phi$, could be used to cool the oscillator to lower temperatures.

\begin{figure*}[!t] \centering
\includegraphics[width=.98\textwidth]{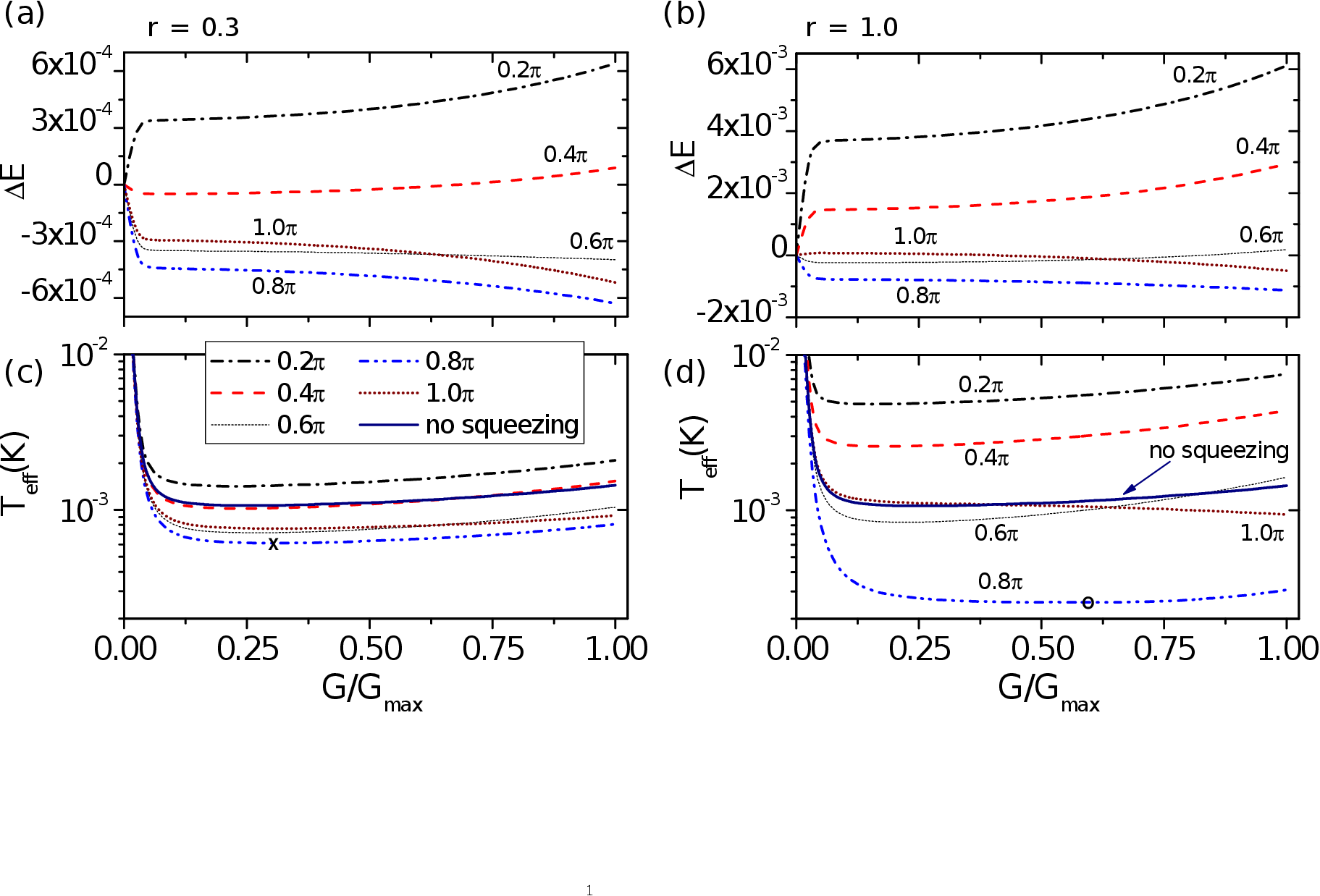}
\caption{(a) and (b) Energy contribution due to squeezing-related terms ($\Delta{E}$) and (c) and (d) the corresponding effective temperature ($T_{eff}$) for $r$ = 0.3 and 1.0, respectively. $\Delta{E}$ could be significantly negative (blue dash-dot-dotted line), causing the oscillator to be deeply cooled (represented by ``x'' and ``o'') below the standard values when no squeezing is performed (navy solid lines).} \label{fig_300}
\end{figure*}
In Fig. \ref{fig_300}(a), the contributions of the terms due to the squeezed field, $\Delta{E}$, in the case $r$ = 0.3, are presented. $\Delta{E}$ changes from positive ($\phi=0.2\pi$, black dashed-dotted line), to nearly zero ($\phi=0.4\pi$, red dashed line), and finally negative ($\phi$ = 0.6$\pi$--1.0$\pi$) values. As a result, the effective temperature ($T_{eff}$, Fig. \ref{fig_300}(c), navy solid line) can be lower than the standard limit value, from the minimal value of $\simeq$ 1 mK to approximately 0.6 mK (represented by ``x'' in the blue dash-dot-dotted line). 

Increasing $r$ to 1.0, $\Delta{E}$ results in a  more negative value [see Fig. \ref{fig_300}(b)] and $T_{eff}$ can be much lower compared to the case of $r$ = 0.3. The minimum temperature i 0.2 mK (represented by ``o'' in the blue dash-dot-dotted line in Fig. \ref{fig_300}(d)). Nevertheless, if $r$ is larger, the difference in $N(\omega)$ and $M(\omega)$ becomes smaller ($\frac{N(\omega)}{M(\omega)}\propto\tanh{r}\longrightarrow$ 1 for $r>$ 2), and the effectiveness of cooling is reduced. 

\begin{figure*}[!ht] \centering
\includegraphics[width=0.7\textwidth]{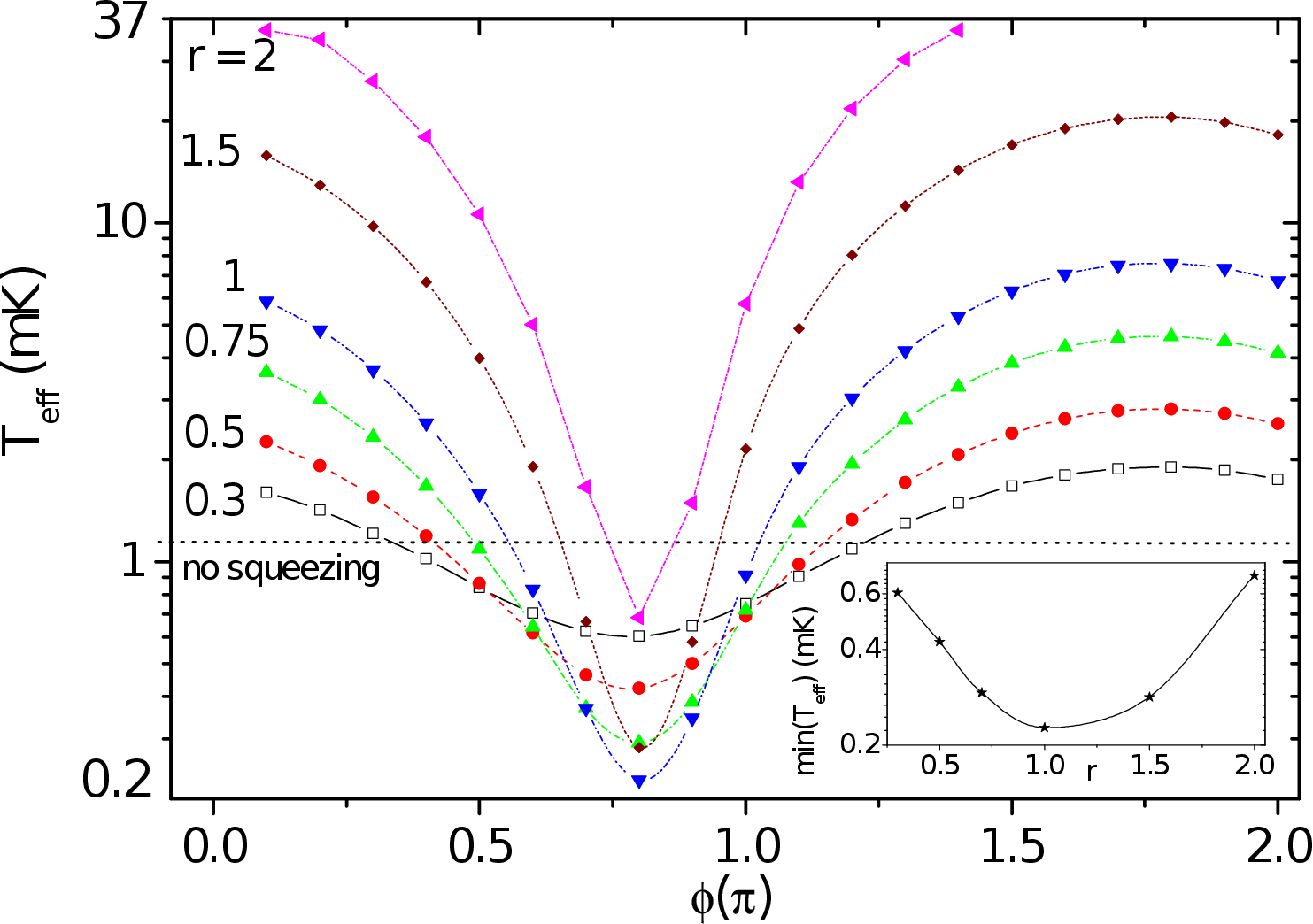}
\caption{Minimum effective temperature obtained using various squeezing parameters, $r$ = 0.3--2, and phases, $\phi$ = 0--2$\pi$. $T_{eff}$ could be as small as 0.2 mK whereas 1 mK is obtained in the case of no squeezing (black horizontal dotted line). (Inset) Summary of minimum $T_{eff}$ versus $r$ (solid line is for guiding the view).} \label{fig_minT}
\end{figure*}
The competition between heating and cooling is mainly due to the change in the sign of $\cos\phi$ and $\sin\phi$ with $\phi$ or more explicitly, of $E_M^c\cos\phi-E_M^s\sin\phi$ in $\Delta{E}$. This term achieves a minimum value at $\phi_M={\arctan}\frac{-E_M^s}{E_M^c}$, for example, $\phi_M(r=1) \simeq$ 0.8$\pi$ (rad). We obtain the lowest effective temperature when $\phi$ is changed from 0 to 2$\pi$ in Fig. \ref{fig_minT}. In this case, the initial temperature of the oscillator is 37 mK. The other parameters are the same as that used in Fig. \ref{fig_Er}. All values of $T_{eff}$ below the dotted line that indicate no squeezing exhibit deep cooling due to the squeezed field; otherwise, they exhibit heating. We could see that for $r=1$, $T_{eff}$ can reach $\simeq$ 0.2 mK (blue dash-dot-dotted line with inverted triangles). As a result, using the suitable squeezing parameter $r$ and phase $\phi$ could cool the oscillator further compared to the standard limit value.
\begin{figure*}[!t] \centering
\includegraphics[width=0.7\textwidth]{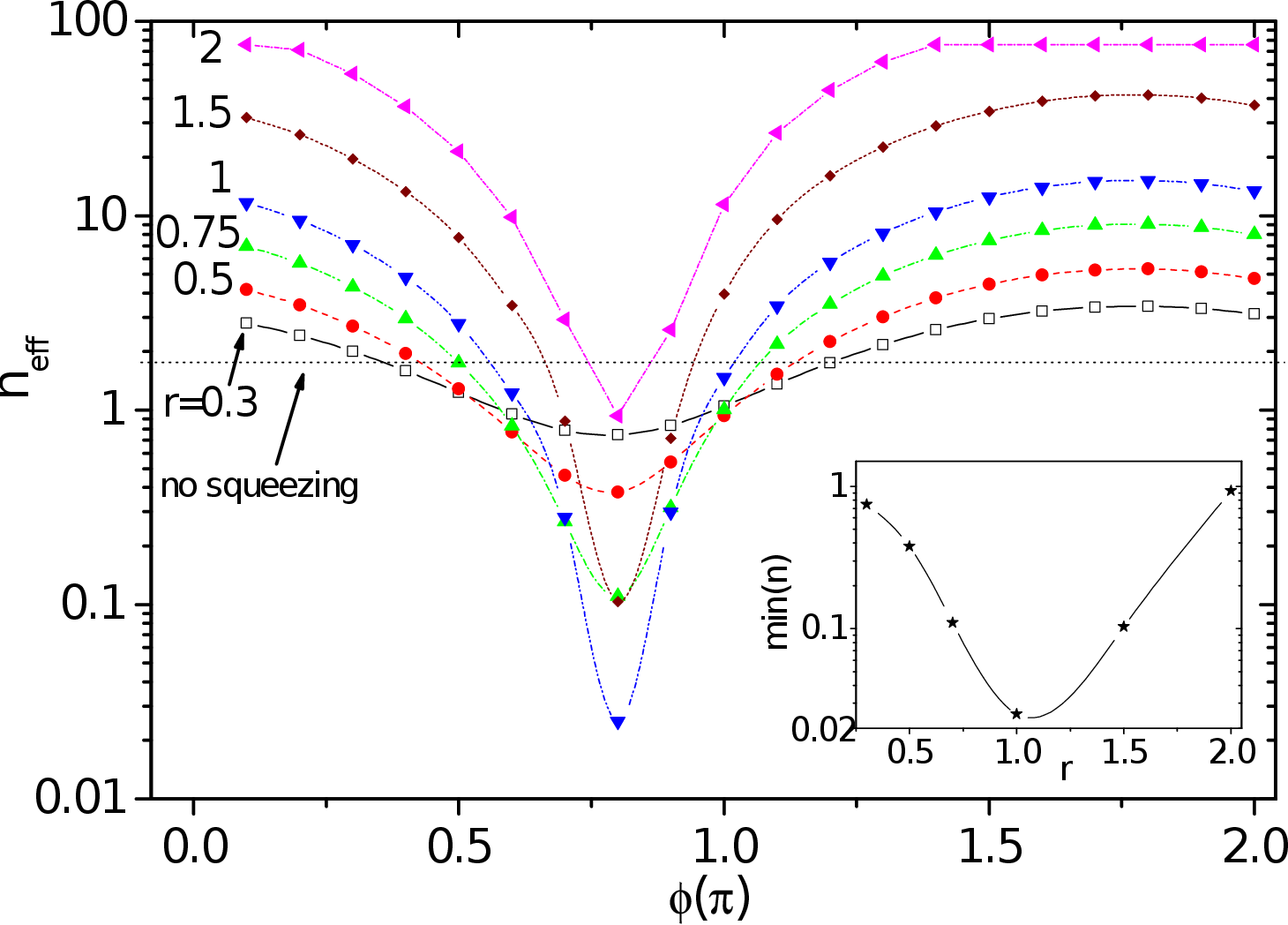}
\caption{Minimum effective quantum number ($n_{eff}$) for various squeezing parameters. $n_{eff}$ could be as small as 0.2 in comparison to 1.7 in the case of no squeezing. (Inset) Summary of minimum $n_{eff}$ versus $r$ (solid line is for guiding the view).} \label{fig_minn}
\end{figure*}

The classical-quantum boundary is usually presented using the effective quantum number $n_{eff}$ of the oscillator, $E_m=\frac{1}{2}\hbar\omega_m(n_{eff}+\frac{1}{2})$. $n_{eff}$ = 0 and 1 correspond to the ground state and the first excited state, respectively. In Fig. \ref{fig_minn}, we present the minimum $n_{eff}$, which corresponds to the effective temperature presented in Fig. \ref{fig_minT}. $n_{eff}$ could be reduced up to three orders of magnitude, from 10.2 to 0.025, by using $r$ = 1 (blue dash-dot-dotted line). It is noteworthy that the initial quantum number, which is determined by the temperature of the thermal bath, should be low enough to achieve $n_{eff}$.

\section{Conclusion}
We have explicitly examined the role of a squeezed field on the heating and cooling of an optomechanical oscillator. Cooling was achieved by choosing appropriate squeezing parameters to create a negative contribution from the field to the oscillator energy. It was shown that a squeezing parameter near unity and a phase of approximately 0.8$\pi$ minimized the effective temperature. A decrease of three orders of magnitude of the temperature could be obtained for optimal cooling. These results highlight important insights on the utilization of squeezed light for the cooling of an oscillator to temperatures close to the classical-quantum boundary.

\section*{Acknowledgments} V.N.T. Pham and N.D. Vy are thankful to Professor V-H. Le (HCMUE) and Assoc. Prof. Tr-D. Ho for encouragement. This research is funded by the Ministry of Education and Training of Vietnam under grant number B2021 - SPS - { }06. 

\bibliographystyle{tMOP}
\bibliography{VybibAPEX}

\end{document}